\documentclass[a4paper,preprintnumbers,floatfix,superscriptaddress,pra,twocolumn,showpacs,notitlepage,longbibliography]{revtex4-1}
\usepackage[utf8]{inputenc}
\usepackage[T1]{fontenc}
\usepackage[sc,osf]{mathpazo}
\usepackage{mathtools}
\usepackage{amsfonts}
\usepackage{bbm} 
\usepackage{booktabs}
\usepackage{xcolor}
\usepackage{diagbox}
\usepackage{float}

\begin{document}

\title{A portrait of the collaboration network in quantum information}
\author{Samuraí Brito}
\affiliation{Emerging Technology Group at Itau-Unibanco}
\affiliation{International Institute of Physics, Federal University of Rio Grande do Norte, 59078-970, Natal, Brazil}
\author{Rute Oliveira}
\affiliation{Federal University of Rio Grande do Norte, Departamento de F\'isica Te\'orica e Experimental, Natal-RN, 59078-900, Brazil.}
\author{Raabe Oliveira}
\affiliation{Federal University of Rio Grande do Norte, Departamento de F\'isica Te\'orica e Experimental, Natal-RN, 59078-900, Brazil.}
\author{Rafael Chaves}
\affiliation{International Institute of Physics, Federal University of Rio Grande do Norte, 59078-970, Natal, Brazil}
\affiliation{School of Science and Technology, Federal University of Rio Grande do Norte, Natal, Brazil}
\date{\today}

\begin{abstract}
From its inception in the beginning of the eighties, with milestone results and ideas such as quantum simulation, the no-cloning theorem and quantum computers, quantum information has established itself over the next decades, being nowadays a fast developing field at the interface between fundamental science and a variety of promising technologies. In this work we aim to offer a portrait of this dynamical field, analyzing the statistical properties of the network of collaborations among its researchers. Using the quant-ph section from the arXiv as our database, we draw several conclusions on its properties. In particular, we show that the quantum information network of collaborations displays the small-world property, is very aggregated and assortative, being also in line with Newman's findings as for the presence of hubs and the Lotka's law regarding the average number of publications per author.
\end{abstract}

\maketitle

\section{Introduction}
Although this was certainly not the original intent, social psychologist Stanley Milgram's experiments became cultural icons of the 1960s. A few years after publishing his results on obedience to authority \cite{milgram1963behavioral}, Milgram would become even more famous with an elaborate experiment \cite{milgram1967small} which allowed him to prove that the urban legend of the ''six degrees of separation`` was a reality. First proposed in the 1920s in a short story by the Hungarian writer Frigyes Karinthy \cite{karinthy2011chain}, this theory implied that six friendship bonds would be enough for two people to be connected. In Karinthy's words: “To demonstrate that people on Earth today are much closer than
ever, a member of the group suggested a test. He offered a bet that we could name any person among earth’s one and a half billion inhabitants and through at most five acquaintances, one of which he knew personally,
he could link to the chosen one”.

Interestingly, a few years before Milgram's experiment, two mathematicians, Karinthy's compatriots, Paul Erdös and Alfréd Rényi, had proposed a mathematical model of networks that naturally led to the six degrees of separation \cite{erdos1959random,erdos1960evolution}. In the Erdös-Rényi's theory all nodes in the network are egalitarian, having the same probability of connecting to each other. Despite its completely random character, the Erdös-Rényi probabilistic network gives rise to the property that the average shortest  path between any two nodes is proportional to the logarithm of the total number of nodes in the network, the famous small-world property fantasized by Karinthy and first observed by Milgram.

Over time, however, it was realized that several other properties observed in natural networks could not be well described by the Erdös-Rényi model. For instance, the clustering coefficient, measuring the connectivity among neighbours of given node of the network, is typically large in the various types of real networks  but it is very small in the Erdös-Rényi approach, a feature that was first reproduced by the Watts-Strogatz model \cite{watts1998collective}. But even their model failed to reproduce several other characteristics that started to be observed and cataloged in the most varied networks \cite{barabasi2016network,barabasi1999emergence,barabasi2002evolution,Barabasi2002,Jeong2000}. An essential characteristic of a network is its connectivity, that is, how much network nodes are connected to each other. And both the Erdös-Rényi and the Watts-Strogatz models predicted that the connectivity is described by a Poissonian, implying that the vast majority of nodes should have a connectivity very close to an average value. Notwithstanding, real networks are not unbiased, they contain the presence of hubs, nodes that concentrate a large part of the connections and play a fundamental role in the global interconnectivity of the network.

\begin{figure*}[t!]
\begin{center}
\includegraphics[scale=.45]{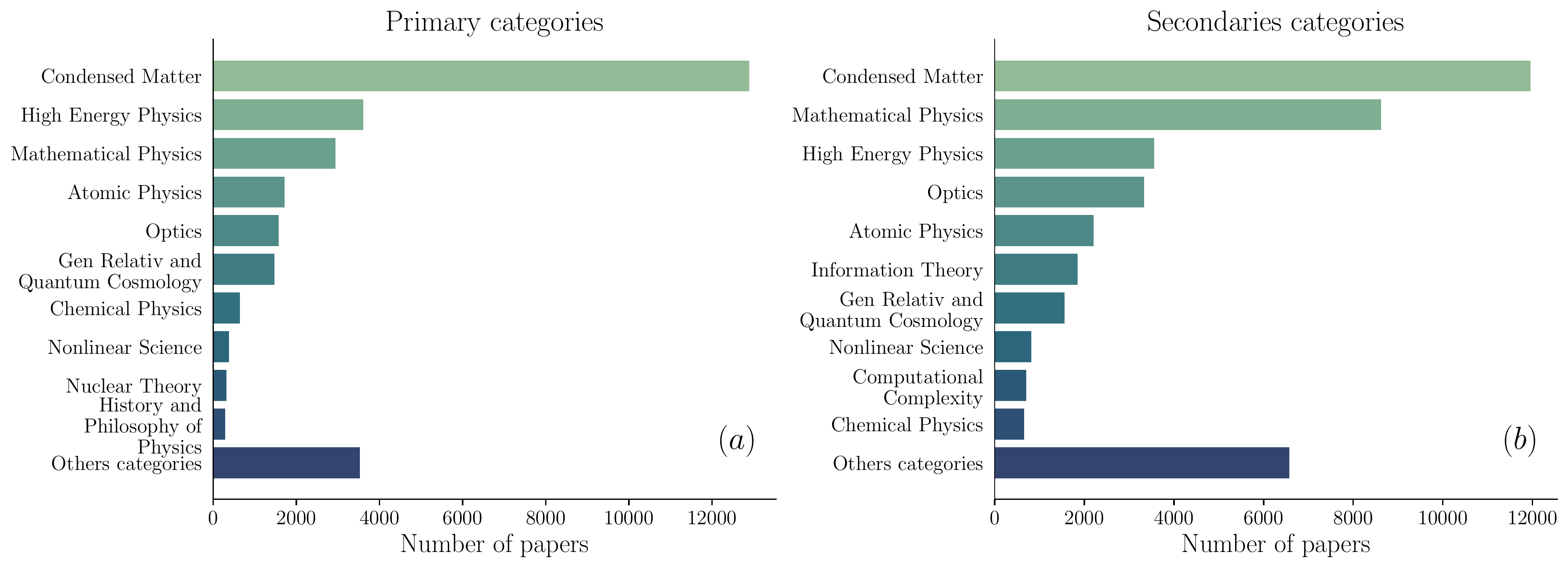}
\end{center}
\caption{ {\bf Top $10$ fields listed as cross-lists with quant-ph.} In $(a)$ we show the number of papers that have another area as primary and \textit{quant-ph} as secondary category and in $(b)$ we show the number of papers that have \textit{quant-ph} as primary category and at least one cross-list area. The last bars in secondaries categories include all other fields of the arXiv appearing as cross-lists with \emph{quant-ph} and that include Cryptography and Security, Computational Physics, Nuclear Theory, Machine Learning, among others.}
\label{cross-listed}
\end{figure*}

A paradigmatic example is the World Wide Web (WWW) network whose connectivity distribution is described by a power-law \cite{albert1999diameter}. Unlike a Poissonian distribution, in which the chance of node deviating from the average degree is exponentially small, in a network ruled by a power law there is no such characteristic average connectivity.
Most nodes have low connectivity but the probability of finding high connectivities decay polynomially rather exponentially.
Reason why such networks are called scale-free. To explain this new property, the Barabási-Albert model \cite{barabasi1999emergence} introduced a new feature called preferential attachment, implying that a new node added to the network is more likely to connect to highly connected nodes than those with few connections.

Over the years network science \cite{barabasi2016network} established itself as a interdisciplinary field offering a common language to study statistical and collective properties of the most varied phenomena, including biological \cite{mason2007graph,meyers2005network}, financial \cite{boginski2006mining,Petrone2018,Caccioli2018}, technological \cite{albert1999diameter,gao2016universal} among many other networks \cite{amara2011classes, samurai1, brito2021satellite, dehmamy2018structural, gysi2021network}. What matters from a network perspective is how connected the constituents of a system are, an approach that led to the realization that networks with a very different nature can behave similarly from a statistical point of view.

Of particular relevance to us, is the fact that network science can be successfully employed to model and analyze a wide range of social networks, in particular scientific networks \cite{barabasi2002evolution}, citations among scientific articles \cite{aydinoglu2018origins,porter2009interdisciplinary,brazelton2009understanding,fay2015scientometric,rosvall2010mapping,seskir2021landscape} as well as the collaboration network of scientists in various fields \cite{newman2001structure, newman2001best, newman2004coauthorship, sun2015we, kumar2014relationship}.

Within this context, our aim here is to analyze the properties of the collaboration network of scientists working in the field of quantum information. Starting with seminal results in the eighties, such as Feynmans's quantum simulation ideas \cite{feynman2018simulating} and the quantum version of Church-Turing thesis proposed by Deutsch  \cite{deutsch1985quantum}, the field has rapidly evolved in the next decade to become one of the most active and fast increasing fields of research nowadays. Arguably, this steadily growing attention is specially due to the promises of a variety of quantum technologies, ranging from efficient quantum simulations \cite{lloyd1996universal} and quantum computation \cite{nielsen2002quantum} to quantum sensors \cite{degen2017quantum}, quantum communication \cite{gisin2007quantum} and cryptography \cite{gisin2002quantum}. Being almost three decades old and formed by scientists that have embraced the arXiv (an open-access repository of electronic preprints) from early on, the quantum information community offers an ideal study case for network science.

The paper is organized as follows. In Sec. \ref{sec:sec2} we describe our methodology to use the \emph{quant-ph} field of the arXiv to construct the network of researchers in quantum information. In Sec. \ref{sec:sec3} we analyze the quantitative aspect of the quantum information community such as the temporal evolution of number of authors and papers. In Sec. \ref{sec:sec4} we focus on the statistical properties of the network allowing us to prove, for instance, the small-world behaviour of this collaboration network. In Sec. \ref{sec:sec5} we show some instances of the network of collaborations of some authors. In Sec. \ref{sec:sec6} we discuss our findings and point out interesting directions for future investigations.

\begin{figure*}[t!]
\begin{center}
\includegraphics[scale=.5]{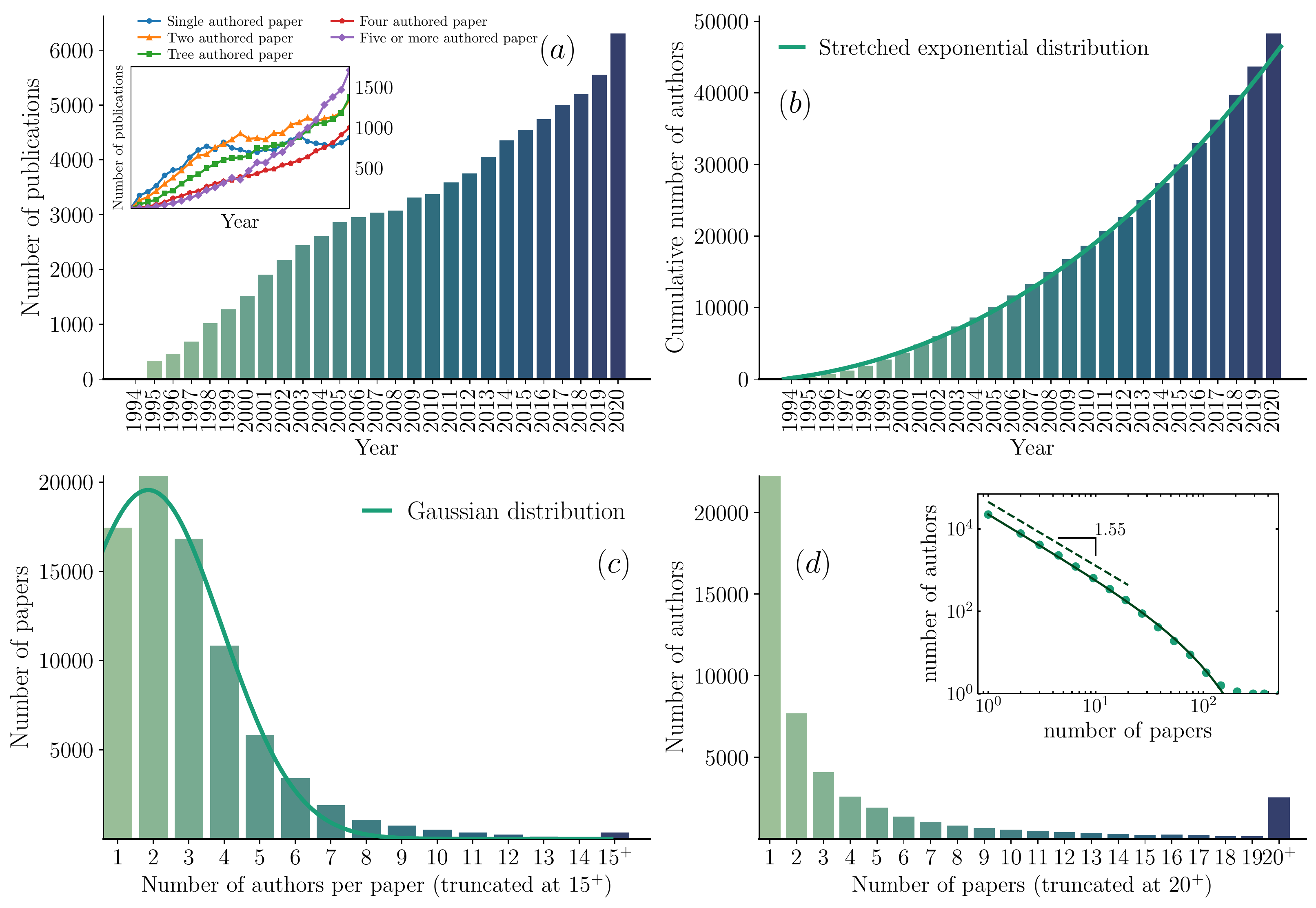}
\end{center}
\caption{ {\bf Results and analysis of the number of articles, authors, authors per papers and author productivity.} $(a)$ Number of papers published per year. The inset shows the number of papers published by a single author, two authors and so on until five or more authors per paper. $(b)$ Cumulative number of authors added per year on the network from 1994 to December 31 of 2020. The data can be fitted by a stretched exponential $(c)$ Histogram of the number of authors per paper, well fitted as a gaussian distribution. $(d)$ Histogram of the number of publications per author. The inset shows the log-log representation of the same data and we use the Lotka's law of scientific productivity with a exponential cutoff given by Eq.~\ref{lotka_law} to fit our data.}
\label{n_pub_and_authors}
\end{figure*}

\section{Constructing the network of scientific collaborations in quantum information science}
\label{sec:sec2}
The first problem one faces when analyzing the scientific networks of a given field is the construction of the database. A typical approach \cite{aydinoglu2018origins,porter2009interdisciplinary,brazelton2009understanding,fay2015scientometric,rosvall2010mapping,seskir2021landscape} is to search scientific databases such as the Web of Science using a pre-defined set of keywords. This, however, might lead to the inclusion of entries that are not representative of the field and the exclusion of certainly relevant works. Arguably, that is particularly prominent in a interdisciplinary field such as quantum information. A query based on ``entanglement'' is likely to include articles and authors from various disciplines within physics, ranging from condensed matter and quantum field theory to cosmology as well as different areas such as neuroscience and even psychology. Here instead, we follow a different approach an use the quant-ph category of the arXiv as a characteristic database of the research in quantum information. Indeed, this community has historically being one of the most active in the arXiv, actively posting their works in this pre-print repository since its inception. As a matter of fact,  milestone results, such as Shor's algorithm \cite{shor1994algorithms} that have boosted the field, are included in that list. One could argue, however, that the quant-ph category also includes research outside the scope of quantum information.
As we will see, while that is certainly true, the network of scientists we construct following this approach generates a giant cluster that subsumes over $87.5\%$ of the networks nodes. That is, the quant-ph can indeed be seen as a faithful portrait of a highly interconnected network, that we assume is a faithful representation of the quantum information research network.

\begin{figure*}[t!]
\begin{center}
\includegraphics[scale=.41]{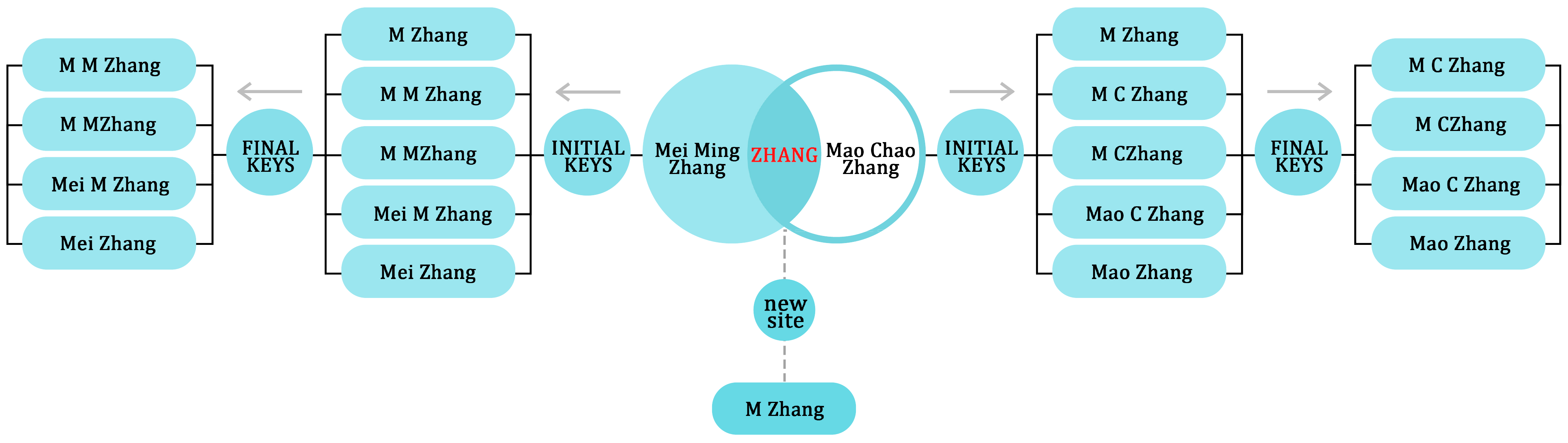}
\end{center}
\caption{\textbf{Examples of different variations of two authors names and keys.} For example, the author \textit{Mei Ming Zhang} at first has the initial keys: \textit{M Zhang}, \textit{M M Zhang}, \textit{M MZhang}, \textit{Mei M Zhang}, and \textit{Mei Zhang}. Comparing these keys with another author, for instance, \textit{Mao Chao Zhang} that has the keys: \textit{M Zhang}, \textit{M C Zhang}, \textit{M CZhang}, \textit{Mao M Zhang}, and \textit{Mao Zhang}, both authors initially have the same key \textit{M Zhang}, therefore is not possible to know which of the authors belongs the paper written by \textit{M Zhang}, because of that we added a new site to assigned to this key, as denote in the figure. In our data there are $504$ authors with surname \textit{Zhang} and $13$ authors with \textit{M} as first initial name and surname \textit{Zhang}. As a result, we identify these cases and remove the key formed by the surname and the first initial only. The surnames of the authors that appear most in the network are: Wang, Zhang, Li, Chen and Liu, with over 350 repetitions each.}
\label{example_keys}
\end{figure*}

Within  this context, we have analyzed the scientific collaboration network of the quantum information community based on the dataset available on the arXiv from $1994$ to December 31 of 2020. We use the R package \textit{aRxiv}~\cite{packageaRxiv} which retrieves the metadata from papers and stores its information, including title, authors, date and category. Considering only those where \emph{quant-ph} appears as one of the categories, the data set is composed of $109534$ entries. However, through our analysis we will focus only on those articles with quant-ph as the primary category, a set of $80165$ papers published in aforementioned timeframe. From those, a total of $53980$ publications have quant-ph as the only category. The remaining $26185$ papers have at least one secondary category and as shown in Fig. \ref{cross-listed}, condensed matter is the field with the largest overlap with quantum information. In turn, the number of papers from other disciplines that had quant-ph as a secondary category is given by $29369$, once more with condensed matter on the top of the intersection with quantum information research.

With the papers data at hand, the first step to create the network of researchers in quantum information was to identify all the different authors and their equivalent names in the data. Once the same author can employ different variations of their name, for instance, \emph{Alice Qubit} or  \emph{A. Qubit} we have grouped these different versions and created a key, consisting of the name with the largest number of characters, for each author.  Hence, it should be noted that there may exist mistakes in distinguishing some authors, especially for those with the same family name and the same initial for the first name (see Fig.~\ref{example_keys}). The final set of authors names define the nodes of the quantum information scientific collaboration network and two authors are linked if they have co-authored a paper. After the pre-processing of the data available at the arXiv, the constructed network has a size of $N=48327$ nodes (authors) with $E=282193$ links (common papers) among them. We have employed Python libraries for complex network research, namely \textit{igraph} and \textit{networkx}, to create and analyze all network properties discussed in the following.

\section{Quantitative aspects of the quantum information community}
\label{sec:sec3}
In this section we focus on the quantitative aspects such as number of papers and researchers, number of co-authors per paper and number of papers per author, also analyzing their time evolution. 

In Fig. \ref{n_pub_and_authors}a we plot the number of papers published per year \footnote{It should be noticed that there is a discrepancy between the data available at the arXiv and its search engine. For instance, while the database at \emph{https://arxiv.org/year/quant-ph/94} retrieves a total of $12$ publications during $1994$, a direct search would return only $11$ entries. A similar issue occurs for every year, with the search engine retrieving less papers than that available at the database (the one we employ to construct our network).}. It becomes clear from the graph, the rising interest quantum information has attracted over the last $25$ years, with an increasing number of papers appearing consistently year after year. In $1994$, the first year of our temporal series, mere $12$ articles have been published whereas in the next year already this number increases to $335$ papers. It is curious to notice that in spite of the COVID-$19$ pandemic starting in $2020$, this was the year with most articles, a total of $6309$. Fig. \ref{n_pub_and_authors}a also shows, as an inset, the number of papers published by a single author and  collaboratively written. It is interesting to notice that until $2005$ most of the papers were made by single authors and that after this date the number of this kind of publications reaches a plateau. In parallel, we observe a clear increase in the number of papers with co-authors, in particular those with five or more authors. Overall, it is reasonable to argue that with the establishment of quantum information as a research area, collaborations among researchers became more common rather than working in isolation. 

\begin{figure*}
\begin{center}
\includegraphics[scale=.7]{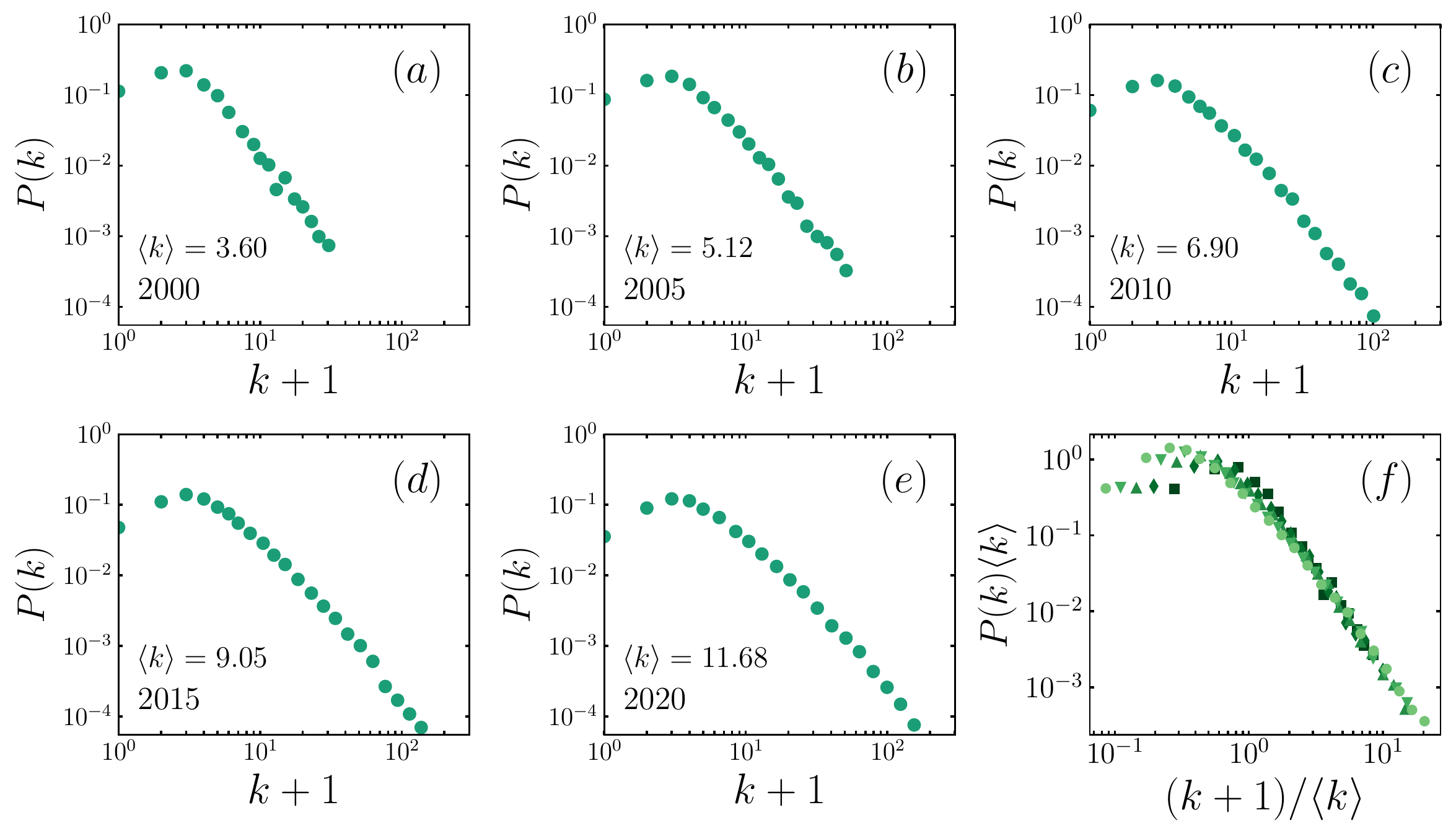}
\end{center}
\caption{ {\bf Degree distribution of the cumulative network of given year, considering a period of $20$ years, separated by  five-year intervals.} Authors who have never collaborated have degree zero, because of this we use the $k + 1$ in the abscissa. $(a)$ In the year $2000$ the network size was $N = 3720$ and contained $E = 6699$ links, with an average degree $\langle k \rangle = 3.60$. $(b)$ In $2005$ the total number of sites in the network increased to $N = 10116$ with $E = 25890$ and $\langle k \rangle = 5.12$. $(c)$ The plot shows the degree distribution in $2010$ with $N = 18637$ sites and $E = 64324$ while $\langle k \rangle = 6.90$. $(d)$ In $2015$ the network evolved to $N = 30009$ with $E = 135865$ edges and $\langle k \rangle = 9.05$. $(e)$ The plot corresponds to the cumulative network in $2020$ with $N =  48327$ sites and $E = 282193$ links and average degree $\langle k \rangle = 11.68$. In $(f)$ the data collapse onto a single curve for the same data in $(a)$ to $(e)$ when expressed in terms of $k / \langle k \rangle$ and rescaling $P(k)\to P(k) \langle k \rangle$. All the degree distributions were logarithmic binned.}
\label{pdk_year}
\end{figure*}

In Fig. \ref{n_pub_and_authors}b we show the cumulative number of authors added per year on the network. The cumulative number of authors grows like a stretched exponential distribution over the years being well fit by
\begin{equation}
    f(t) = A e^{\alpha(t - t_0)^{\beta}}
\end{equation}
where $A$ is a normalizing constant and $\alpha = 5.21$, $t_0 = 1994$ and $\beta = 0.2$ are constants.

In turn, Fig. \ref{n_pub_and_authors}c shows the total number of papers with a given number of co-authors. Most of the papers contain two authors but a significant fraction of papers contain single or three authors. In contrast, papers with $15$ or more authors are much more rare, a total of $362$ entries. In our dataset we could identify only two papers displaying the phenomenon known as hyperauthorship \cite{cronin2001hyperauthorship}, scientific articles with a disproportional large number of authors (beyond $100$), corresponding to references \cite{yu2020quantum} and \cite{big2018challenging}. Not surprisingly, those entries correspond to large scientific collaborations, the LIGO and the Big Bell Test, respectively. As also shown in Fig. \ref{n_pub_and_authors}c, the histogram of the number of authors per paper is well fitted by a Gaussian distribution where the probability density function generic form is given by
\begin{equation}
p(x) = A e^{-\frac{1}{2}\left(\frac{x - \mu}{\sigma}\right)^2},
\end{equation}
where the parameter $\mu = 0.858$ is the mean, the parameter $\sigma = 2.08$ is the standard deviation and A is a normalization constant.

Interestingly, as shown in Fig. \ref{n_pub_and_authors}d, we observe that most of the authors in the field have authored only one article. This is likely due to the fact that quantum information has seen an increasing number of students that, while still being in the beginning of their careers, have only published a single article with their advisors or research groups. On the other hand, the most productive author in our dataset has published $523$ papers.

It is interesting to compare our data with the result established by Lotka in $1926$ that the distribution of scientific production of researchers is described by a power-law~\cite{lotka1926frequency}. It describes the frequency with which scientists in a certain field get published. In his analysis, however, just senior authors were taken into account. Later on, Newman suggested an exponential correction in the Lotka's law ~\cite{newman2001best} that naturally appeared due the finite size effects. Our results are consistent with the Lotka's law of scientific productivity with a exponential cutoff,
\begin{equation}
    f(X) \propto X^{-n} e^{-l_p X}
    \label{lotka_law}
\end{equation}
where $X$ is the number of publications, $f(X)$ is the proportion of authors who have $X$ publications, $l_p$ is the characteristic cutoff length and $n$ depends on the specific field of knowledge. Using this equation to fit our data we obtain the constants $n = 1.55$ and $l_p = 65$ (see Fig. \ref{n_pub_and_authors}d). The cutoff length is connected to the exponential decay of the curve; the lower the value of this parameter, the quicker the curve decays, and as a result, the maximum number of publications of an author is smaller. For comparison, we note that for fields such as the biomedicine \cite{newman2004best} and of Brazilian researchers on the Lattes platform \cite{araujo2014collaboration} the values of the power law exponent are given by $n=2.86$ and $n=1.58$, respectively. This exponent suggest that the quantum information community has a sizable disparity where there are few researchers who have a high publish rate and many researchers who publish much less.
For convenience, all data used to generate the plots in  Fig. \ref{n_pub_and_authors} is available as tables in the Appendix.

\section{Statistical Properties of the Network}
\label{sec:sec4}
A central goal of network science is to understand the statistical properties as well as the asymptotic behaviour of networks as the number of network nodes increases. Mathematically, a network is defined as a graph $G = \{V, E\}$, where $V$ is a set of $N$ nodes and $E$ are the connections (edges) among the elements of $V$ \footnote{We highlight that to create our network we are excluding the two papers mentioned before that display the phenomenon of hyperauthorship. We observe that the inclusion of these papers introduce significant fluctuations for the assortativity (to be defined below) but apart from that all other statistical properties remain mostly unaltered.}. All the most important statistical quantities of our network are listed in Table \ref{prop_net}. In particular, we notice that the network has a total of $3086$ clusters, that is, sub-networks that are not interconnected. As expected, however, the giant cluster comprises $87.5 \%$ of all nodes, with the second largest cluster containing only $28$ nodes and being composed of a small research group from other fields such as Mathematical Physics, Nuclear Theory and Condensed Matter. As paradigmatic in network science, the average shortest path and the diameter are calculated over the giant cluster (otherwise, by their own definition they would amount to infinite for nodes in disconnected sub-networks). All other quantities refer to the whole network.

Within this context, a central quantity in the understanding of networks its their degree distribution $P(k)$, the probability of finding a node with k degree. For the Erd\"os and R\'enyi (ER) model based on random graphs \cite{erdos1959random, erdos1960evolution}, for sufficiently large $N$, the degree distribution $P(k)$ can be approximated by the Poissonian $P(k) = \frac{e^{-\langle k \rangle} \langle k \rangle^k}{k!}$, where $\langle k \rangle=p(N-1)$ is the average connectivity of the network and $p$ is the probability that a nodes connects with another (a constant for all nodes of the network). In turn, many real networks contain nodes with a high degree of connections, the hubs that are described by the preferential attachment mechanism of the Barabási–Albert model generating random scale-free networks \cite{barabasi1999emergence}.

\begin{table}
\centering
\caption{Statistical properties of the quant-ph network} 
\label{prop_net}
\begin{tabular}{p{18em} p{8em}} 
 \hline
 Total papers & $109534$ \\ 
 Total papers with  quant-ph  & $80165$ \\
 as the primary category  & \\
 Total papers with quant-ph  & $29369$ \\
 as the secondary category  & \\
 Total authors & $N = 48327$ \\
 Total collaborations & $E = 282193$ \\
 Mean authors per paper & $3.16$ \\
 Mean papers per author & $5.23$ \\
 Number of clusters & $3086$\\
 Size of giant cluster& $N_G/N = 0.875$\\
 Second largest cluster & $28$ nodes\\
 Assortativity coefficient & $r = 0.139$\\
 Clustering coefficient & $\langle C \rangle = 0.646$\\
 Shortest path (minimum distance) & $\langle l \rangle = 4.73$\\
 Diameter (maximum distance) & $d = 18$\\
 Collaborators per author & $\langle k \rangle=11.68$\\
 Most connected author & $k = 660$\\[1ex] 
 \hline
\end{tabular}
\end{table}

We analyze the temporal evolution of the network connectivity over the years. In Fig. \ref{pdk_year} we display the connectivity distribution over a period of $20$ years, considering five years intervals. Every year, new authors join the network increasing the number of nodes. As  well, new collaborations are established also increasing the number of links in the network. Visually, the distributions are very similar, yet the proportions grows each year.

As shown Fig.~\ref{pdk_year}f, the quantum information collaboration network, has a skewed degree distribution~\cite{faloutsos2011power} where just a few scientists act as hubs and have a lot of connections while the bulk of scientists have only a few collaborations. This result is in line with Newman's finding that the degree distribution does not completely follow a power-law shape, but rather has an exponential cutoff \cite{newman2001structure}. It is worthy noting that this type of degree distributions is also observed on other collaboration networks as in neuroscience \cite{barabasi2002evolution} and earth sciences \cite{kumar2014relationship}. As can be seen in Table \ref{deg_rank}, showing the ten most connect authors, quantum information science has a number of researchers acting as hubs that interconnect a significant fraction of networks, including authors who have worked with over $650$ collaborators.

\begin{figure}
\begin{center}
\includegraphics[scale=.8]{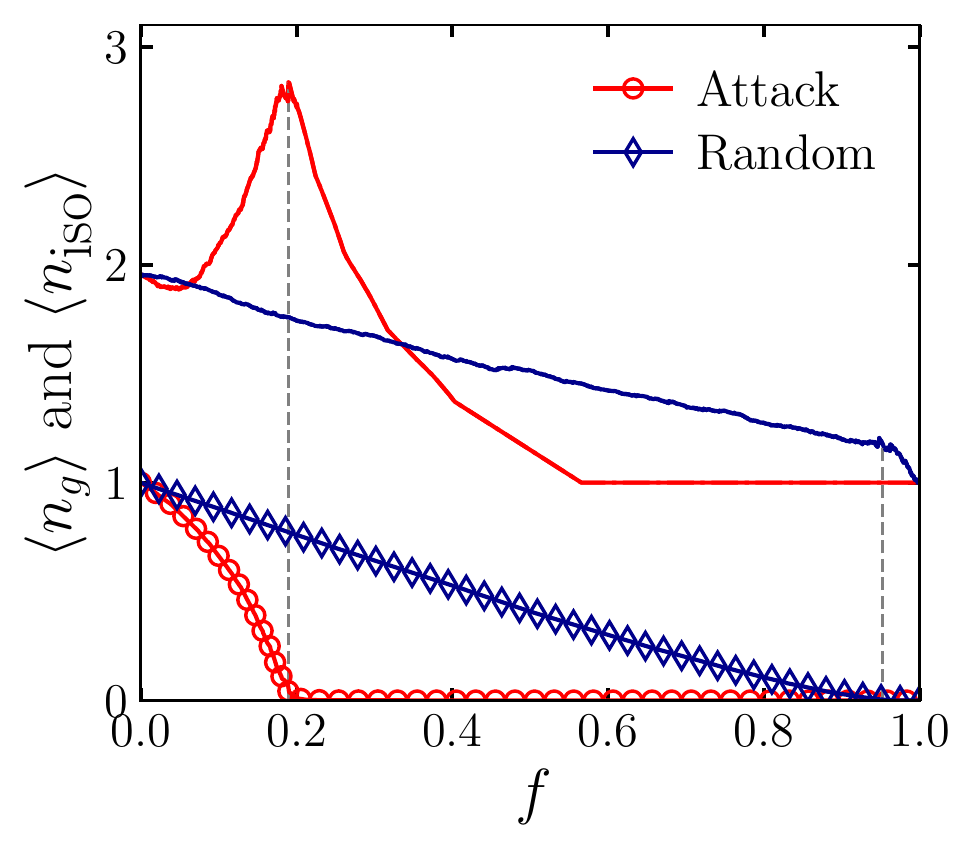}
\end{center}
\caption{ {\bf Quantum information collaboration network robustness.} In this figure we compare $\langle n_g \rangle$ (red circles and blue diamonds) with $\langle n_{\mbox{iso}}\rangle$ (red straight line and blue dashed-dotted line) as a function of the fraction $f$ of removed nodes. The network initially has $3086$ isolated clusters at first, with only $1719$ them being formed by a single-site cluster, this implying that the average size of the isolate clusters is more than $1$ before we start to remove the sites. We see that the network is robust against random failures and less robust against targeted attacks.}
\label{robustness}
\end{figure}

Motivated by the presence of these hubs we performed the analysis of network robustness. Our objective is to figure out how many nodes have to be removed from the network for it to break down.  We consider both random failures (a random node is removed from the network in each interaction) and targeted attacks (the most connected nodes of the network are removed sequentially). As shown in Fig.~\ref{robustness}, we follow the standard procedure in the literature \cite{albert2000error}, analyzing the ratio $\langle n_g \rangle = N_G(f)/N_G(0)$ and $\langle n_{\mbox{iso}}\rangle$ as the fraction of the removed nodes by the network size $f$, where $N_G(f)$ is the size of the giant cluster after we remove a fraction of nodes, $N_G(0)$ is the size of the initial giant cluster and $\langle n_{\mbox{iso}}\rangle$ is the average size of the isolate clusters without the giant cluster. As can be seen, in targeted attacks it is necessary to remove $18.9\%$ of the sites to bring down the network, while under random  failures almost all of the nodes of the network, more exactly $95.2\%$, need to be removed to disconnect the network. The results are compatible with scale free networks that are robust against random failures and fragile against target attacks \cite{barabasi2016network}. An interesting fact is that the network have $3086$ clusters initially,  $1719$ composed of authors who have never co-authored a paper and $1367$ clusters with authors that collaborate in small groups plus the giant cluster. Consequently, the average size of the isolate clusters is initially bigger than $1$ (not including the size of the giant cluster, of course).

\begin{figure}
\begin{center}
\includegraphics[scale=.57]{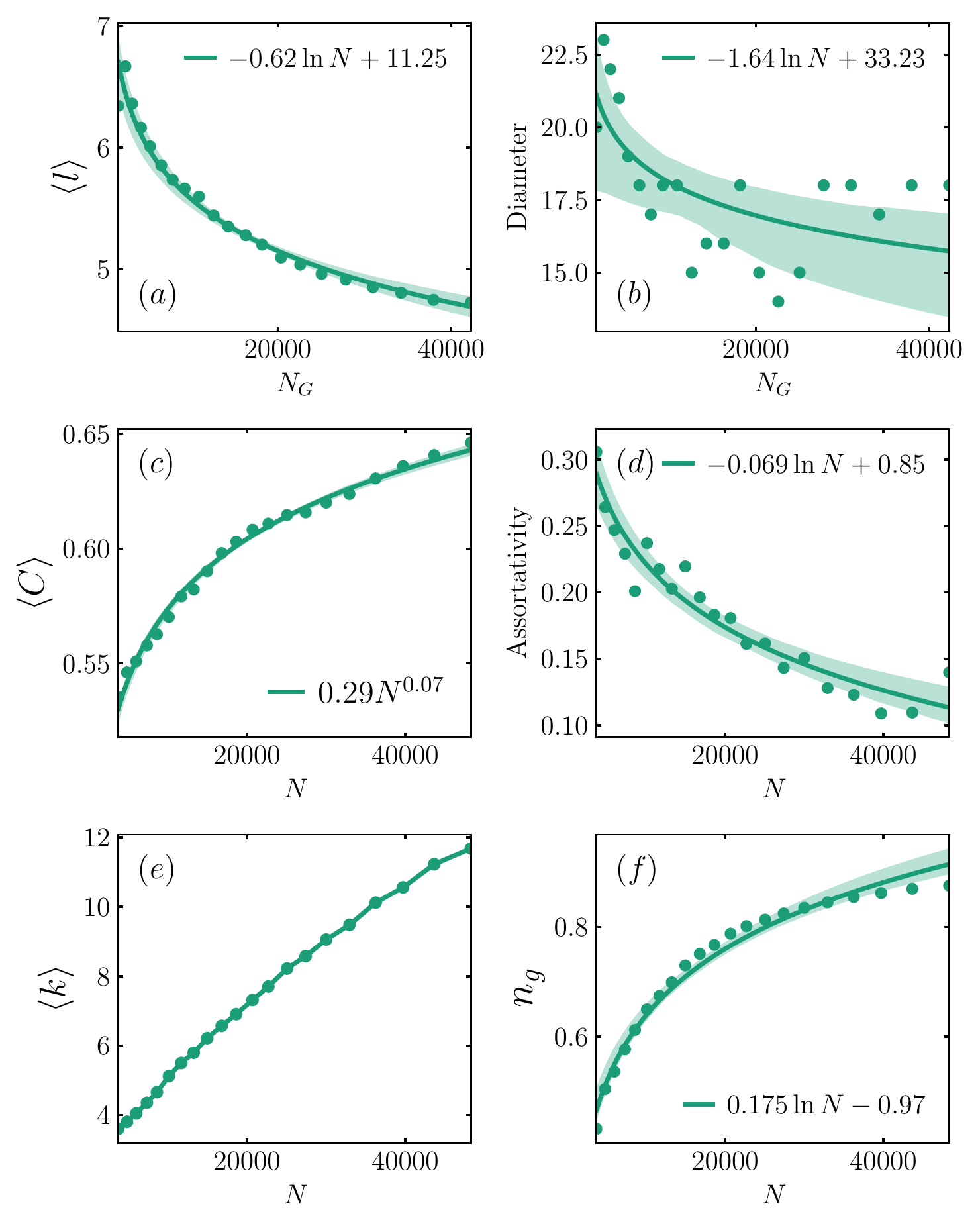}
\end{center}
\caption{ {\bf Some properties in the study of networks.}  In $2001$ the network size was $N = 4821$ and $2020$ goes to $N = 48327$. $(a)$ average path length, $(b)$ diameter, $(c)$ clustering coefficient, $(d)$ assortativity of the network, $(e)$ average degree and $(f)$ relative size of giant cluster $n_g = N_G/N$. In all plots we inset a regression line to guide our eyes.}
\label{prop-py}
\end{figure}

We also analyzed the temporal evolution of the average degree $\langle k \rangle$, the average number of links per node quantifying how many collaborators each author has. To guide our eyes we connect the points in the figure and we can observe a logarithmic tendency on this property, as shown in Fig. \ref{prop-py}e. However we can observe that $\langle k \rangle$ growth with the network's size $N$ and that can be explained by the typical fact that the number of collaborations of a scientist grows over the years. At the end of our time series, corresponding to the end of $2020$, the average connectivity of authors was $\langle k \rangle=11.68$.

Typically, any two sites in the network will be connected by more than one path. However, more often we are interested in the shortest path among them, the minimum number of links required to connect any two sites. This notion of distance in a network is encompassed by  the average shortest path length $\langle l \rangle$  defined as $\langle l \rangle = \frac{2}{N(N-1)} \sum_{i < j} l_{i j}$, where $l_{i j}$ is the shortest path between nodes $i$ and $j$. If there is no edge between two sites $i$ and $j$ the distance is defined as $l_{i j} = \infty$, reason why we have to refer to the largest cluster of the network.

The phenomenon known as \textit{small world} refers exactly to the fact that the average shortest path length $\langle l \rangle$ between two nodes scales logarithmic with the network size. For instance, for random networks $\langle l \rangle = \ln N / \ln \langle k \rangle$. As showed in the Fig. \ref{prop-py}a, the shortest path in our network decreases logarithmically as the size of the giant cluster increases over time, being very well fitted as $\langle l \rangle = -0.62\ln N+11.25$. That is, the quantum information network has the small-world property.

\begin{table}
\centering
\caption{Alphabetical order of the ten most connected authors with the corresponding average clustering coefficient $\langle C \rangle$ of the first neighborhood network for each of them. We also compute the local clustering coefficient of each listed author defined as $c_{\text{local}} = \frac{2n_i}{k_i(k_i - 1)}$, 
where $n_i$ is the number of edges
between the $k_i$ neighbors of the site $i$ and $k_i(k_i - 1)/2$
is total possible number of edges between them. Note that the local clustering coefficient of each author is significantly smaller than $\langle C \rangle$, meaning that each author have small groups that the participants strongly collaborate between them, increasing $\langle C \rangle$. The number of connections (collaborations) of these authors varies between $395$ to $660$.} 
\label{deg_rank}
\begin{tabular}{p{13em} p{6.5em} p{6.5em}} 
 \hline
Author & $\langle C \rangle$ & $ c_{\text{local}} $\\ [0.5ex] 
\hline
Anton Zeilinger & $0.767$ & $0.072$ \\
Franco Nori & $0.737$ & $0.022$\\
Guang Can Guo & $0.755$ & $0.024$ \\ 
Ian Alexander Walmsley & $0.740$ & $0.058$ \\
Jian Wei Pan & $0.731$ & $ 0.051 $\\
Martin Bodo Plenio & $0.665$ & $ 0.029 $\\
Mikhail D. Lukin & $0.733$ & $ 0.044 $\\
Nicolas Gisin & $0.682$ & $ 0.045 $\\
Sae Woo Nam & $0.760$ & $ 0.057 $\\
Vlatko Vedral & $0.695$ & $ 0.044 $\\[0.5ex] 
 \hline
\end{tabular}
\end{table}

Related to the average shortest path is the diameter of the network, the maximum  shortest path between any two nodes of the network. As can be seen in Fig. \ref{prop-py}b, most likely due the size of the network, we still observe significant fluctuations in the diameter from year to year and the logarithmic regression we employ should only be seen as visual guide to compare with the data. It is clear, however, a decreasing trend in the diameter of the network over time. At the end the time series in $2020$, the diameter was given by $d = 18$, slightly smaller when compared to the diameter of the condensed matter collaboration network found to be $d=22$ around $20$ years ago \cite{newman2004best}.

Another relevant property for the statistical analysis of networks is the \textit{average clustering coefficient}  $\langle C \rangle = \frac{1}{N}\sum_{i} c_i$, where $c_i = \frac{2n_i}{k_i(k_i - 1)}$ is the local clustering coefficient of the site $i$ and
$n_i$ is the number of edges between the $k_i$ neighbours of the site $i$ and $k_i(k_i - 1)/2$ is total possible number of edges between them. Within our context, it can be understood as a measure how much the collaborators of a given author tend to also collaborate among themselves. For random graphs, for instance, this coefficient is typically very small and given by $\langle C \rangle =  \langle k \rangle / N $ (thus decreasing with the size of the network). As can be seen in Fig. \ref{prop-py}c, the average clustering coefficient is increasing over time for the quantum information network, being well fitted as $\langle C \rangle=0.29 N^{0.07}$. By the end of 2020, the quantum information community had a considerably high average clustering coefficient of $\langle C \rangle=0.646$ indicating that the researchers tend to establish a close and interlinked network of collaborations. This is close to the clustering coefficient of the research collaboration in computer intelligence in games  \cite{lara2014analysis} that, similarly to the quantum information case, has increased over the years. In contrast, the biomedical research collaboration network \cite{newman2001structure}  has a low clustering coefficient, $\langle C \rangle = 0.066$, implying that collaborators of  a scientist in biological research are far less likely to have papers in common.

We also analyze the assortativity coefficient, $-1\leq r\leq 1$, reflecting the nodes affinity for linking with other nodes displaying some sort of similarity. If $r>0$, the network is  considered assortative with examples such as student relationships and the network of actors and actresses \cite{newman2002assortative}, while the world-wide web \cite{newman2003mixing} and protein interactions are examples of disassortative networks \cite{newman2003mixing}. Here we measure the assortativity coefficient with respect to the node's degree, a measure also known as  Pearson's correlation coefficient and given by 
\begin{equation}
    r = \frac{M^{-1}\sum_i j_i k_i - \left[M^{-1} \sum_i \frac{1}{2}(j_i + k_i)\right]^2}{M^{-1} \sum_i \frac{1}{2}(j_i^2 + k_i^2) - \left[ M^{-1} \sum_i \frac{1}{2}(j_i + k_i) \right]^2}
\end{equation}
where $j_i$, $k_i$ are the degrees of the sites at the end of the $i$th link, with $i = 1, ..., M$ and $M$ is the total number of links in the network. 

As can be seen in Fig. \ref{prop-py}d, the quantum information network begins as a highly assortative. The assortativity shows significant fluctuations, an expected behaviour for a new field of research where the sub-areas of expertise are not yet defined. However, as the number of researchers increases over time and the field matures, we see clearly that the network always remains  assortative, a trend similar to that observed in other research fields in physics as whole \cite{newman2002assortative} as well as in information and library science \cite{sun2015we}.

Finally, we also study the network's giant cluster evolution. As can be seen in see Fig. \ref{prop-py}f, the size of the largest cluster in the network increases over time, starting around $50\%$ and reaching over $87\%$ in the end of 2020. Due to the new collaborations established by researchers over the years as well as the new partnerships formed with new authors, not only the number of isolated clusters decreases but also the relative size of the largest cluster increases over time.

\section{Visualizing the network}
\label{sec:sec5}

Another relevant network property is the community structure~\cite{girvan2002community}. Communities are qualitatively defined by nodes that are densely interconnected while the connections between different communities tend to be sparser. In our scenario, a community can be interpreted as research group or a collection of research groups that have close interactions. To determine the communities in our networks we use \emph{Modularity}, an algorithm for community detection by Blondel \textit{et al}.~\cite{blondel2008fast} implemented in \textit{Gephi}~\cite{bastian2009gephi}. This method (also called 
Louvain method) is based on a local optimization of the Newman-Girvan modularity~\cite{newman2004finding} for the general case of weighted networks. The quantity to be optimized is called modularity, defined for an unweighted network (like our case) as
\begin{equation}
\label{eq:modularity}
    Q = \frac{1}{2m} \sum_{ij} \left( A_{ij} - \gamma\frac{k_ik_j}{2m}\right)
    \delta(c_i,c_j)
\end{equation}
where $m$ is the number of links, $A$ is the adjacency matrix of the network, $k_i$ is the degree of site $i$, $\gamma$ is a resolution parameter, $c_i$ is the community to which the $i$-th node of the network is assigned and $\delta(c_i,c_j)$ is the Kronecker delta function, $1$ if $i=j$ and $0$ otherwise. The higher is $\gamma$ the bigger are the resulting communities within the whole network and vice-versa. We set $\gamma = 1$ by default and this value return a network with a maximum of $7$ communities.

The optimization of the modularity follows two sequential steps. The first step assumes that each site is assigned to a different community, with the number of communities initially equal to the size of the network. Then, given a site $i$ the algorithm computes the increase in the modularity (Eq.~\ref{eq:modularity}) if $i$ is removed from its community and placed in one of the communities of its neighbors. If there is no gain, the node stays in its original community, otherwise is moved to the community where the increase in the modularity is maximum. The communities generated in this first step are called supervertices, and two supervertices are connected if there is at least one link between sites of the corresponding communities. In the second step, the exchange of communities of a given site continues in order to maximize the modularity but now considering exchanges not among neighbours but regarding the supervertices.  Clearly, the number of communities on the network can only decrease with each interaction, unless the modularity achieves a maximum and constant value, in which case the optimization ends.

\begin{figure*}[ht!]
\begin{center}
\includegraphics[scale=1]{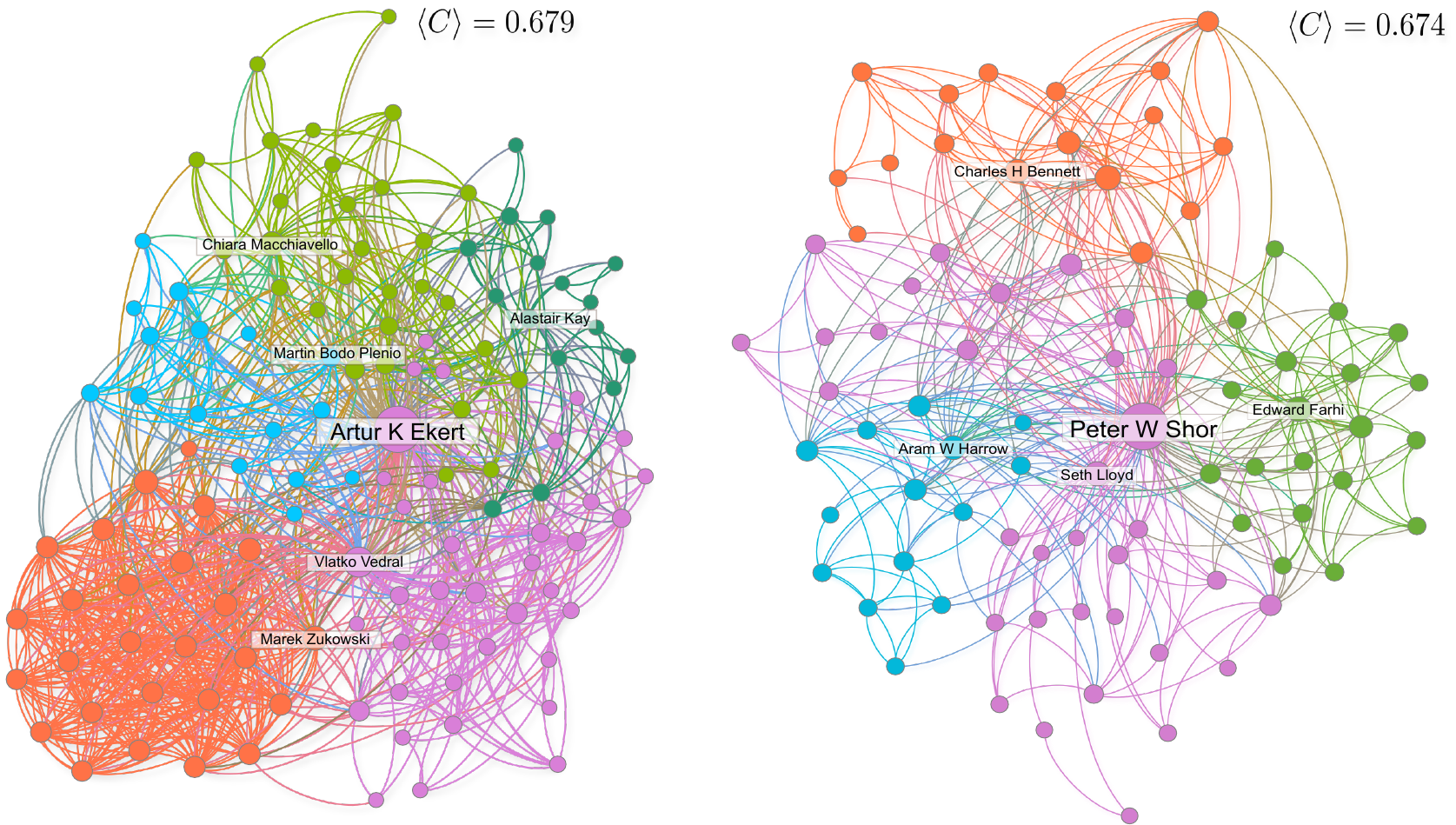}
\end{center}
\caption{\textbf{Examples of collaboration networks} On the left the network of Artur K. Ekert and on the left the network of Peter W. Shor. The first network is composed of $N = 126$ authors with a total $E = 957$ links, an average degree of $\langle k \rangle \simeq 15$ and clustering coefficient $\langle C \rangle =0.679$, presenting $5$ different communities. Similarly, for Shor we obtain $N = 85$,  $E = 400$, $\langle k \rangle \simeq 9$ and $\langle C \rangle =0.674$ and a total of $4$ communities. To facilitate the visualization, we display only the names of the  authors with higher degree within a given community.}
\label{network_gephi}
\end{figure*}

Examples of authors networks are shown in Fig.~\ref{network_gephi}. The colors indicate different communities and the labels are shown only for authors with the higher degree in a given community. It can be visually seen that members of a given community/research group frequently cooperate with one another, but just a few authors from these communities/groups collaborate with authors from other communities/groups. The choice for Artur K. Ekert and Peter W. Shor was made for two main reasons. They are authors of great relevance for the field but with a moderate number of collaborations, so that their whole network may be visualized accurately. Furthermore, even though their research interests certainly overlap, Ekert most cited work regards quantum cryptography while Shor's is about a quantum algorithm. In this sense, they represent different sub-areas within quantum information. In spite of that, we see that their collaboration networks show quite similar clustering coefficients, $\langle C \rangle = 0.679$ for against  $\langle  C \rangle = 0.674$. In turn, other network properties differ. For instance, we obtain $N=126$ co-authors for Ekert and $85$ for Shor, a total of $E=957$ links in Ekert's network and $400$ in Shor's, and an average degree $\langle k \rangle \simeq 15$ and $\langle k \rangle \simeq 9$, respectively.

\section{Discussion}
\label{sec:sec6}
Quantum information is a relatively new and multidisciplinary field at the interface between physics, computer science and information theory, among others. Given the promises of quantum technologies, this research area has attracted a steadily increasing attention. Using tools and concepts from network science, the central goal of this work is to make a portrait of this dynamic and very active area of research. For that, we have constructed the collaboration network of researchers in quantum information, using the \emph{quant-ph} database from arXiv spanning the time period from $1994$ until the end of $2020$.

First, focusing on the quantitative aspects of this network, the data shows that the number of publications as well as the number of researchers in the area have seen a significant increase over the years. From a modest number of $300$ authors in $1995$, the year Peter Shor published his milestone result \cite{shor1994algorithms}, the field had by the end of $2020$ over $48.000$ researchers, an expansion that is well described by an exponential increase. As we show, the number of papers per author is compatible with the Lotka's law of scientific productivity \cite{lotka1926frequency} and the number of researchers per paper is governed by a Gaussian distribution centered around two authors per paper. It is curious to notice that as the years passed by, the number of authors per paper has changed from a a majority of single authored articles to publications involving collaborations with five or more co-authors, a clear sign that quantum information has matured as an interdisciplinary research field where partnerships are important.  

Regarding the statistical properties of the collaboration network, we have shown that the connectivity distribution of authors is governed by a skewed degree distribution: the majority of researchers have few connections while a few scientists act as hubs of the network, a result in line with Newman's findings \cite{newman2001structure}. Analyzing the time evolution of a few quantities of interest we could show that the quantum information network displays the small-world property, since the average shortest path between two nodes in the network scales logarithmically with the network size, being given by $ \langle l \rangle=4.73$ at the end of 2020. In turn, the diameter of the network, the maximum shortest path, fluctuates considerably but presents a clear sign of decrease as time passes by, being given by $\langle d \rangle = 18$ by the end of our time series. Interestingly, the average connectivity shows a linear growth in time, and by the end of $2020$, a researcher in quantum information had, on average, $\langle k \rangle= 11.68$ collaborators. The clustering coefficient of the quantum information community, measuring how much the collaborators of a given scientist collaborate among themselves, is considerably high, having increased over the years up to $\langle C \rangle=0.646$. The assortativity of the network fluctuates over time but always remain positive and with $r>0.1$, showing that researchers in quantum information tend to collaborate with other scientists that a have similar degree of connections. The size of the giant cluster of the network has also increased with time, being over $87\%$ nowadays, thus showing that \emph{quant-ph} is a highly interconnected  network. Finally, we have also analyze the robustness of this collaboration network, showing under targeted attacks (sequential removal of the most connected nodes), a removal of $18.9\%$ is enough to break down the network, while with the removal of random nodes this number increases to $95.2 \%$.

Our work provides a broad overview of the researchers working in quantum information and we hope it might be trigger further analysis. For instance, it would be interesting to analyze the affiliation, both the universities and research institutes as well as the countries of the authors in the network. Most of the collaborations are local or involve partnerships among different countries and institutions? How is the mobility of researchers in quantum information? How often do they change their affiliation and country? Other relevant direction would be to understand the different subareas within quantum information and how they relate to each other. Unfortunately, however, all this information is not available as metadata in the arXiv submissions and would require the analysis of more informative databases such as the Web of Science or similar services. A first step in this direction has been done in Ref. \cite{seskir2021landscape} but with a focus on the literature rather on the researchers working on quantum information an related fields. 
It could also be interesting to make a similar analysis we have done here but focusing on specific countries or geographic areas, a study that certainly can be use as the basis for establishment and development of national and international quantum technologies research programs.
In the future, we hope to be provide an online platform where the data is constantly updated also providing the tools for everyone to analyse the quantum information network, create and visualize their own network of collaborators (see Fig.~\ref{network_gephi}).

\section{Acknowledgements} 
This work was supported by The John Templeton Foundation via the grant Q-CAUSAL No. $61084$ (the opinions expressed in this publication are those of the author(s) and do not necessarily reflect the views of the John Templeton Foundation), by the Serrapilheira Institute (Grant No. Serra-$1708$-$15763$), the Simons Foundation (Grant Number 884966, AF), the Brazilian National Council for Scientific and Technological Development (CNPq) via the National Institute for Science and Technology on Quantum Information (INCT-IQ) and 406574/2018-9 and 307295/2020-6, the Brazilian agencies MCTIC, CAPES and MEC.

\bibliography{main}

\section{Appendix}
Below we provide tables with the detailed data and information employed to generate the plots in Fig.~\ref{n_pub_and_authors}.

\begin{table*}
\centering
\caption{Number of articles, cumulative authors and authors per papers} 
\label{fig1_ab}
\begin{tabular}{p{5em} p{4.5em} p{5.5em} p{5.5em} p{5.5em} p{5.5em} p{7em} p{8em}} 
 \hline
Year & Number of papers & Single authored & Two authored & Three authored & Four authored & Five or more authored & Cumulative number of authors\\ [0.5ex] 
 \hline
$ 1994 $ & $ 12 $ & $ 6 $ & $ 2 $ & $ 1 $ & $ 3 $ & $ 0 $ & $ 22 $ \\
$ 1995 $ & $ 335 $ & $ 161 $ & $ 100 $ & $ 50 $ & $ 17 $ & $ 7 $ & $ 368 $ \\
$ 1996 $ & $ 463 $ & $ 206 $ & $ 144 $ & $ 79 $ & $ 23 $ & $ 11 $ & $ 700 $ \\
$ 1997 $ & $ 688 $ & $ 281 $ & $ 217 $ & $ 113 $ & $ 49 $ & $ 28 $ & $ 1197 $ \\
$ 1998 $ & $ 1021 $ & $ 410 $ & $ 305 $ & $ 185 $ & $ 77 $ & $ 44 $ & $ 1889 $ \\
$ 1999 $ & $ 1275 $ & $ 476 $ & $ 381 $ & $ 224 $ & $ 127 $ & $ 67 $ & $ 2729 $ \\
$ 2000 $ & $ 1519 $ & $ 493 $ & $ 470 $ & $ 306 $ & $ 153 $ & $ 97 $ & $ 3720 $ \\
$ 2001 $ & $ 1906 $ & $ 634 $ & $ 563 $ & $ 378 $ & $ 196 $ & $ 135 $ & $ 4821 $ \\
$ 2002 $ & $ 2176 $ & $ 723 $ & $ 652 $ & $ 423 $ & $ 214 $ & $ 164 $ & $ 5988 $ \\
$ 2003 $ & $ 2439 $ & $ 769 $ & $ 671 $ & $ 499 $ & $ 273 $ & $ 227 $ & $ 7340 $ \\
$ 2004 $ & $ 2604 $ & $ 730 $ & $ 756 $ & $ 550 $ & $ 307 $ & $ 261 $ & $ 8607 $ \\
$ 2005 $ & $ 2864 $ & $ 818 $ & $ 794 $ & $ 601 $ & $ 338 $ & $ 313 $ & $ 10116 $ \\
$ 2006 $ & $ 2957 $ & $ 749 $ & $ 852 $ & $ 627 $ & $ 350 $ & $ 379 $ & $ 11690 $ \\
$ 2007 $ & $ 3034 $ & $ 727 $ & $ 926 $ & $ 631 $ & $ 393 $ & $ 357 $ & $ 13261 $ \\
$ 2008 $ & $ 3077 $ & $ 693 $ & $ 862 $ & $ 652 $ & $ 404 $ & $ 466 $ & $ 14962 $ \\
$ 2009 $ & $ 3312 $ & $ 694 $ & $ 870 $ & $ 746 $ & $ 432 $ & $ 570 $ & $ 16791 $ \\
$ 2010 $ & $ 3375 $ & $ 731 $ & $ 853 $ & $ 749 $ & $ 475 $ & $ 567 $ & $ 18637 $ \\
$ 2011 $ & $ 3590 $ & $ 717 $ & $ 933 $ & $ 787 $ & $ 488 $ & $ 665 $ & $ 20693 $ \\
$ 2012 $ & $ 3754 $ & $ 796 $ & $ 933 $ & $ 791 $ & $ 537 $ & $ 697 $ & $ 22696 $ \\
$ 2013 $ & $ 4061 $ & $ 845 $ & $ 1034 $ & $ 817 $ & $ 559 $ & $ 806 $ & $ 25059 $ \\
$ 2014 $ & $ 4354 $ & $ 905 $ & $ 1063 $ & $ 884 $ & $ 595 $ & $ 907 $ & $ 27409 $ \\
$ 2015 $ & $ 4549 $ & $ 829 $ & $ 1120 $ & $ 963 $ & $ 638 $ & $ 999 $ & $ 30009 $ \\
$ 2016 $ & $ 4742 $ & $ 806 $ & $ 1076 $ & $ 1053 $ & $ 713 $ & $ 1094 $ & $ 32953 $ \\
$ 2017 $ & $ 4995 $ & $ 789 $ & $ 1112 $ & $ 1055 $ & $ 756 $ & $ 1283 $ & $ 36268 $ \\
$ 2018 $ & $ 5198 $ & $ 774 $ & $ 1131 $ & $ 1103 $ & $ 813 $ & $ 1377 $ & $ 39730 $ \\
$ 2019 $ & $ 5556 $ & $ 813 $ & $ 1187 $ & $ 1179 $ & $ 910 $ & $ 1467 $ & $ 43671 $ \\
$ 2020 $ & $ 6309 $ & $ 876 $ & $ 1345 $ & $ 1378 $ & $ 1001 $ & $ 1709 $ & $ 48327 $ \\[1ex] 
 \hline
\end{tabular}
\end{table*}

\begin{table}
\centering
\caption{Numbers of authors per paper} 
\label{fig1_c}
\begin{tabular}{p{15em} l} 
 \hline
Numbers of authors & Numbers of papers\\ [0.5ex] 
 \hline
$ 1 $ & $ 17451 $ \\
$ 2 $ & $ 20352 $ \\
$ 3 $ & $ 16824 $ \\
$ 4 $ & $ 10841 $ \\
$ 5 $ & $ 5826 $ \\
$ 6 $ & $ 3411 $ \\
$ 7 $ & $ 1905 $ \\
$ 8 $ & $ 1079 $ \\
$ 9 $ & $ 745 $ \\
$ 10 $ & $ 517 $ \\
$ 11 $ & $ 355 $ \\
$ 12 $ & $ 247 $ \\
$ 13 $ & $ 136 $ \\
$ 14 $ & $ 114 $ \\
$ 15^+ $ & $ 362 $ \\[1ex] 
 \hline
\end{tabular}
\end{table}

\begin{table}
\centering
\caption{Number of papers per author} 
\label{fig1_d}
\begin{tabular}{p{15em} l} 
 \hline
Number of papers & Number of authors\\ [0.5ex] 
 \hline
$1$ & $22246$\\
$2$ & $7702$\\
$3$ & $4108$\\
$4$ & $2605$\\
$5$ & $1927$\\
$6$ & $1382$\\
$7$ & $1049$\\
$8$ & $833$\\
$9$ & $663$\\
$10$ & $561$\\
$11$ & $487$\\
$12$ & $418$\\
$13$ & $379$\\
$14$ & $324$\\
$15$ & $254$\\
$16$ & $265$\\
$17$ & $235$\\
$18$ & $174$\\
$19$ & $178$\\
$20^+$ & $2540$ \\
 \hline
\end{tabular}
\end{table}

\end{document}